\documentclass[twocolumn]{aastex631}
\usepackage[utf8]{inputenc}
\usepackage[T1]{fontenc}
\usepackage{amsmath}
\usepackage{amssymb}
\usepackage{mathpazo}
\usepackage{hyperref}
\usepackage{tikz}
\usepackage{bm}
\usepackage{lmodern}
\usepackage[all]{nowidow}
\usetikzlibrary{decorations.pathmorphing}
\tikzset{snake it/.style={decorate, decoration=snake}}
\usepackage{graphicx}
\usepackage{chngcntr}
\usepackage{natbib}
\usepackage{amsmath}
\usepackage[acronym,xindy]{glossaries}
\usepackage{fancyhdr}
\usepackage{environ}
\makeatletter
\newsavebox{\measure@tikzpicture}
\NewEnviron{scaletikzpicturetowidth}[1]{%
  \def\tikz@width{#1}%
  \begin{lrbox}{\measure@tikzpicture}%
  \BODY
  \end{lrbox}%
  \pgfmathparse{#1/\wd\measure@tikzpicture}%
  \BODY
}

\begin{document}

\title{Photon-likeness of hadron showers and impact of Lorentz boosting}


\author{Jannis Pawlowsky}
\affiliation{Bergische Universität Wuppertal, Department of Physics, Wuppertal, Germany}

\author{Karl-Heinz Kampert}
\affiliation{Bergische Universität Wuppertal, Department of Physics, Wuppertal, Germany}

\author{Julian Rautenberg}
\affiliation{Bergische Universität Wuppertal, Department of Physics, Wuppertal, Germany}

\correspondingauthor{\href{mailto:pawlowsky@uni-wuppertal.de}{pawlowsky@uni-wuppertal.de}}

\begin{abstract}
We examine the probability of proton-induced air showers at $E>10$\,EeV being misidentified as photon-induced due to neutral pions receiving a major part of the primary energy in the first interaction, thereby enhancing the electromagnetic shower component by their $\pi^0 \to \gamma \gamma$ decay. Using CORSIKA simulations, we demonstrate the relevance of this effect at EeV energies. However, the probability for such photon-like events drops down strongly at the highest energies due to the increasing probability of Lorentz boosted $\pi^0$'s suffering hadronic interactions before decay. Different hadronic interaction models suggest that photon-like hadronic events may be observed at current UHECR observatories. A quantitative comparison of the observed number of background events found in recent photon searches published by the Pierre Auger Collaboration suggests that the hypothesis of upward fluctuations in $\pi^0$-production alone is insufficient to explain the data. Detailed detector simulations will be needed to quantify the significance of these differences.
\end{abstract}


\section{Introduction}
\begin{figure*}
\centering
\begin{minipage}{0.45\textwidth}
\centering
\includegraphics[width=1\textwidth]{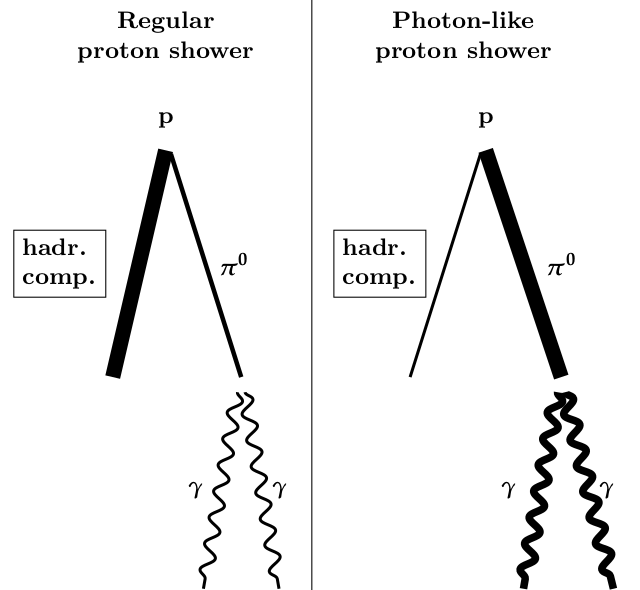}
\caption{Sketch of different energy sharings between charged hadrons and neutral pions. The width of the lines indicates the energy share. Left: Regular proton shower. Right: Photon-like proton air shower}
\label{photon-like}
\end{minipage}%
\hspace{5mm}
\begin{minipage}{0.505\textwidth}
\centering
\includegraphics[width=1\textwidth]{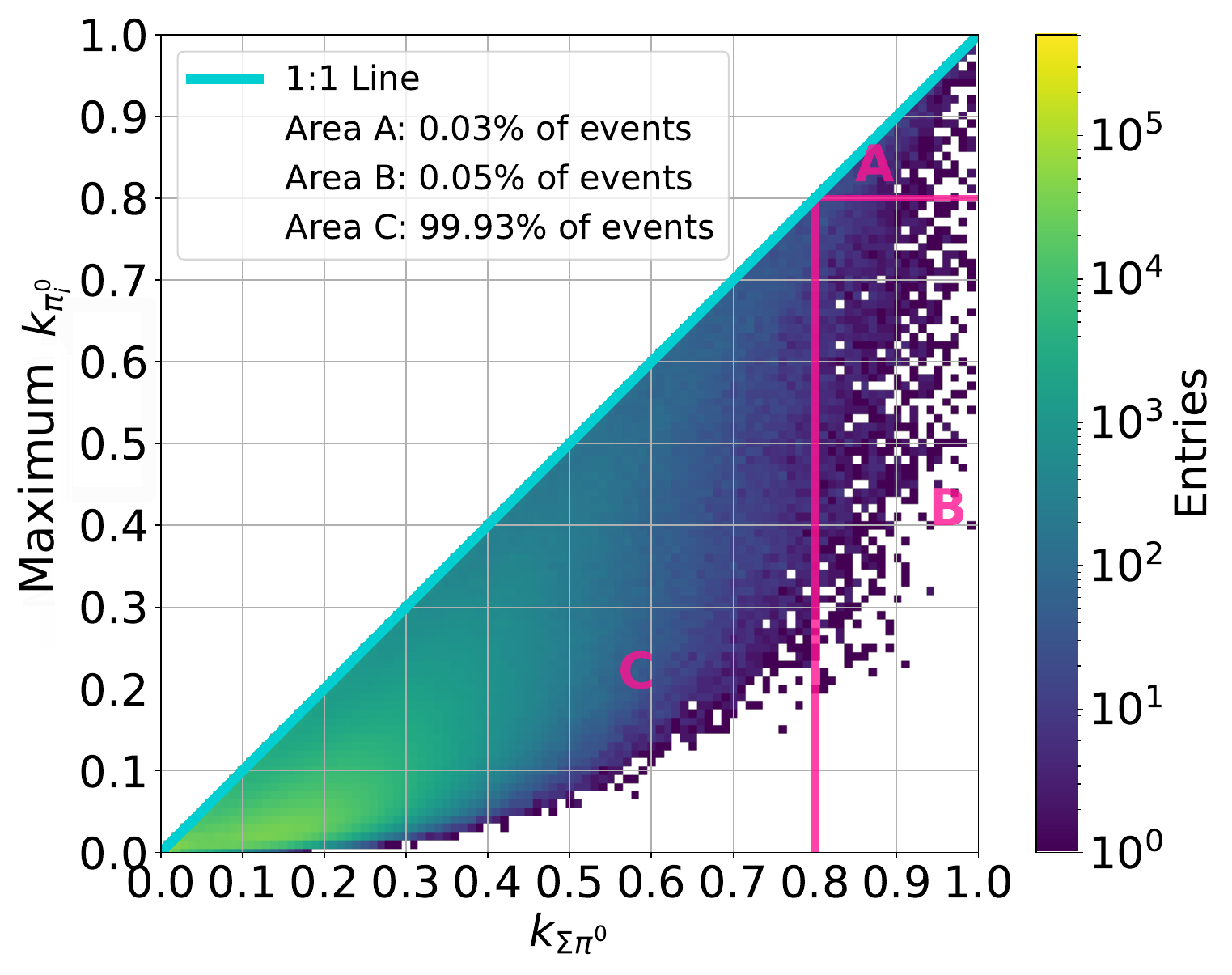}
\caption{Energy fraction of the most energetic $\pi^0$ vs.\ $k_{\Sigma \pi^0}$ for the EPOS-LHC model. The magenta lines mark the areas selected in previous studies (\citealt{lu_thesis}).}
\label{plot_heat_sumvsleading_epos}
\end{minipage}
\end{figure*}

The detection of ultra-high energy (UHE) photons is one of the main objectives in high-energy astroparticle and multi-messenger physics. As their trajectories point directly back to their sources, the detection of photons at $E \gtrsim 0.1$\,EeV would be a major breakthrough in identifying the sources of UHE cosmic rays (UHECRs).
A detection at even higher energies, $E \gtrsim 100$\,EeV, would further open an unexpected window, revealing either new physics or some new particle acceleration mechanisms (\citealt{Addazi:2021xuf,Anchordoqui:2021crl,PierreAuger:2022jyk,PierreAuger:2022ubv,PierreAuger:2023vql}).
Recent studies by the Pierre Auger Collaboration\,(\citealt{PierreAuger:2022aty,PierreAuger:2022gkb}) allowed to set the strongest existing upper limits on the flux of UHE photons for energies $E \geq 10\,$EeV. In that study, 16 UHE photon candidates were identified after applying a Fisher discrimination analysis to 12 years of data. The reported events were found to be consistent with the estimated number of background events so that no detection has been claimed. However, the result motivates further quantitative investigations to verify whether these candidates may represent actual photon-induced events or could originate as background of the much more abundant UHECR induced air showers. 
Amongst UHECR induced air shower, protons penetrate deepest into the atmosphere (\citealt{Kampert:2012mx}) and constitute the main source of background to photon-induced air showers. At the highest energies, the particle showers comprise billions of secondary particles and at each interaction point within the shower, neutral pions can be produced, feeding the electromagnetic component (\citealt{Kampert:2012mx}). Statistically, there is only a small probability that one or multiple highly energetic $\pi^0$'s are produced in the first interaction, but in such cases their decay into photons may contribute significantly to the electromagnetic part of the shower, thereby increasing the photon-likeliness of the air shower. Figure \ref{photon-like} illustrates the situation: in case charged pions receive the largest fraction of the primary energy $E_0$ (l.h.s.), the shower will be hadron-like and vice versa, if neutral pions receive a major fraction of the primary UHECR energy. We note that such charge fluctuations may arise from fluctuations in the initial pp- and pn-collisions in p-nucleus and nucleus-nucleus collisions.  For a quantitative study, we will denote the fraction of primary energy transferred to $\pi^0$'s in the first interaction by 
\begin{align}\label{k_def}
k_{\Sigma \pi^0} =  \sum\limits_{i=1}^{\# \pi^0 } k_{\pi^0,i} = \sum\limits_{i=1}^{\# \pi^0 } \frac{ E_{\pi^0,i}}{E_0},
\end{align}
where the index $i$ represents the different produced $\pi^0$'s and $k_{\Sigma \pi^0} \in [0,\,1]$.

\section{Simulation data set}\label{basics}

This study is based on simulations produced with CORSIKA version 7.7100\,(\citealt{Heck:1998vt}) and FLUKA 2011.2\,(\citealt{Ferrari:2005zk}). As high energy interaction models, QGSJetII-04\,(\citealt{Ostapchenko:2010vb}) and EPOS-LHC\,(\citealt{epos:2013ria}) are employed. Both models are run with the same initial conditions, including azimuth and zenith angles, random seed, and primary energy, as shown in Table \ref{used_geometry}.

\begin{table*}[!htb]
\caption{Overview of simulated cosmic ray interactions}
\label{used_geometry}
\raggedleft
\begin{tabular}{|l|l|l|l|l|l|}
\hline
Property    & $\log_{10}(E/{\rm eV)}$ & Zenith angle [$^\circ$] & Azimuth angle [$^\circ$] & \begin{tabular}[c]{@{}l@{}}Geomagnetic field \\ $B_x/B_z$ [\textmu T]\end{tabular} & \begin{tabular}[c]{@{}l@{}}No.\ of simulated \\ first interactions\end{tabular}               \\ \hline
Value/Range & 19 - 20.5               & 0 - 60       & 0 - 360       & \begin{tabular}[c]{@{}l@{}}19.812/-14.3187 \\ (Auger location)\end{tabular}        & \begin{tabular}[c]{@{}l@{}}5M EPOS-LHC p\\ 200k EPOS-LHC He\\ 2.8M QGSJetII-04 p\end{tabular} \\ \hline
\end{tabular}
\end{table*}

Previous studies examining the background induced by $\pi^0$-production in initial interactions have been presented in Refs.\ \citealt{lu_thesis,Cazon:2020jla}. However, they were limited to investigating the effect of the most energetic $\pi^0$'s. In this study, we account more generally for multiple high energy $\pi^0$'s produced in a single interaction. In fact, it is found more likely that a high energy interaction produces multiple high-energy particles than just a single one. This is illustrated in the heatmap of Fig.\ \ref{plot_heat_sumvsleading_epos}, where for each event the primary energy fraction carried by the most energetic $\pi^0$ is shown as a function of $k_{\Sigma \pi^0}$. A clear peak at a low energy share $k_{\Sigma \pi^0}$ is observed. However, a small proportion of showers is located within the designated areas A and B, and these are the most important areas for this investigation. Previous studies, such as \citealt{lu_thesis} have solely examined events from area A of the plot.  Area B increases the number of analyzed events by a factor of 3. We will argue that events with multiple high energy $\pi^0$'s (Area B) are indistinguishable from events with just one high energy $\pi^0$ (Area A). To illustrate this, we consider the extreme case in which 80\% of the primary energy $E_0$ is distributed over $4\pi^0$s, each receiving 20\% of $E_0$. Such a shower differs from a shower in which a single $\pi^0$ receives 80\% of $E_0$ mostly by developing its shower maximum at two radiation lengths shallower atmospheric depth. This is because the shower with a single $\pi^0$ (Area A in Fig.\,\ref{plot_heat_sumvsleading_epos})  must go through two generations of pair-production and Bremsstrahlung to be comparable to the shower of four initial $\pi^0$'s (Area B in Fig.\,\ref{plot_heat_sumvsleading_epos}). The difference of two radiation lengths lies within the shower-to-shower fluctuations. Thus, showers with multiple initial $\pi^0$-production are indistinguishable on a shower-by-shower basis from showers in which a single $\pi^0$ has taken away a major fraction of the primary energy and need to be included when estimating the hadron-induced background to photon-like showers.

\begin{figure*}[!t]
\centering
\begin{minipage}{0.445\textwidth}
\centering
\includegraphics[width=1\columnwidth]{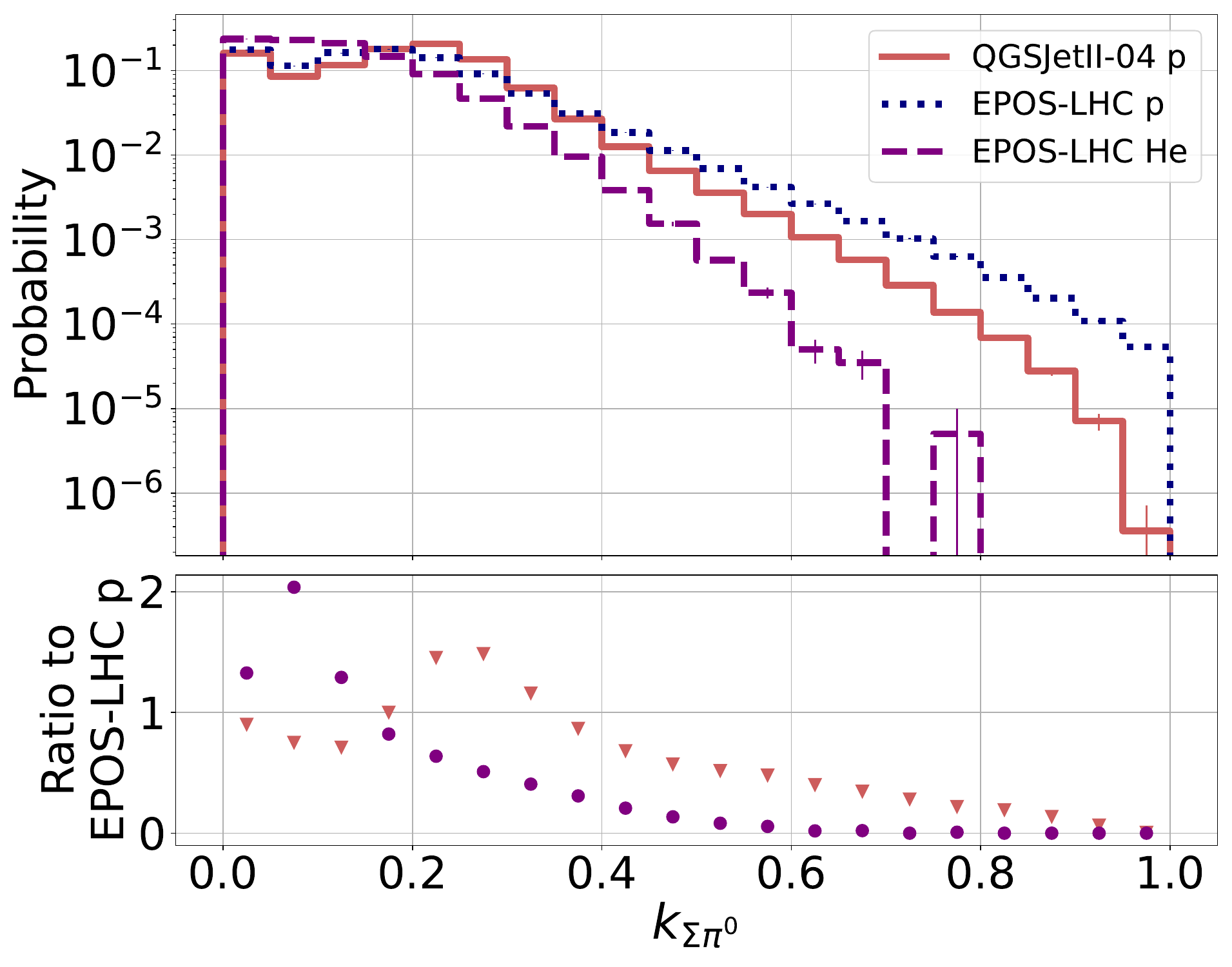}
\caption{Probability distributions of $k_{\Sigma \pi^0}$ for the EPOS-LHC model (blue: proton primary, purple: helium primary) and QGSJetII-04 model (red: proton primary).}
\label{distr_all}
\end{minipage}%
\hspace{5mm}
\begin{minipage}{0.515\textwidth}
\centering
\includegraphics[width=1\columnwidth]{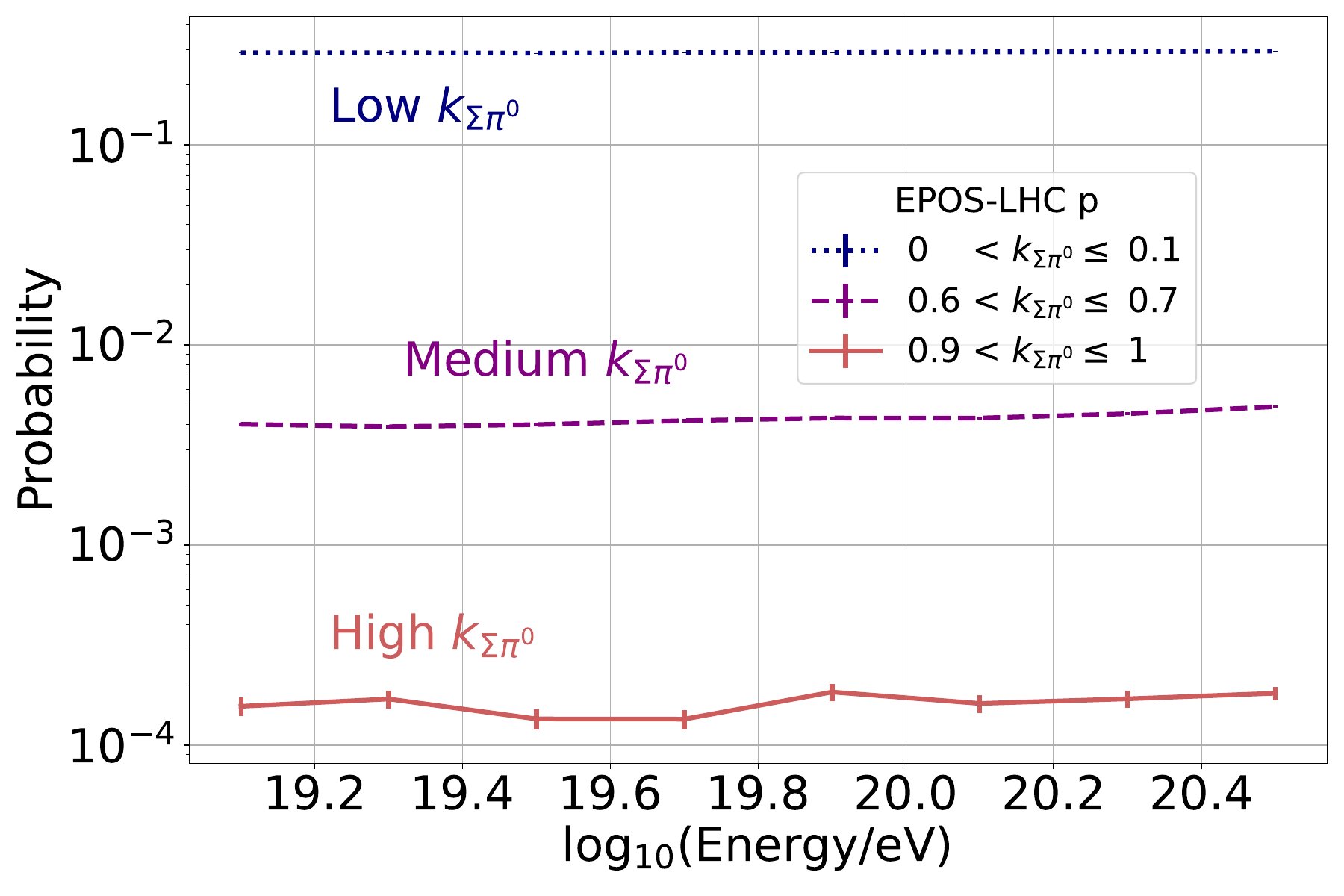}
\caption{Probability to find specific ranges of $k_{\Sigma \pi^0}$ (blue: low, purple: medium, red: large) as a function of the primary energy in proton induced air showers simulated with EPOS-LHC.}
\label{prob_epos}
\end{minipage}
\end{figure*}

\section{Photon-like event probability}\label{chap3}

The probability of finding large values of $k_{\Sigma \pi^0}$ depends primarily on the nature of the primary particle but is also found to depend on the hadronic interaction model used in the simulations. Both aspects are illustrated in Fig.\ \ref{distr_all}. The probability for finding events with $k_{\Sigma \pi^0} \geq 0.7$ in simulations of proton primaries of $E_0 > 10$\,EeV remains below 1\textperthousand, both for EPOS-LHC and QGSJetII-04. When comparing the two interaction models, we find the upward fluctuations of $k_{\Sigma \pi^0}$ to be up to 10-20 times higher in EPOS-LHC as compared to QGSJetII-04.
The impact of the primary particle type is even more significant: Only one event with $k_{\Sigma \pi^0}$ > 0.7 is found amongst $2\times 10^5$ simulated helium events while a few hundred events are observed for the same number of injected proton primaries, regardless of the hadronic interaction model.

To obtain the most conservative estimate for a photon-like background, we will focus on the proton simulations produced with EPOS-LHC and calculate the probabilities for finding specific regions of $k_{\Sigma \pi^0}$ as a function of the primary energy (\textit{c.f.} Fig.\ \ref{prob_epos}). The probability related  to the $\pi^0$ with the most energy after the first interaction is similar to those reported in Ref.\ \citealt{lu_thesis} but are slightly lower when comparing the values of $k_{\pi^0}$. This difference could be related to the newer versions of the models used in this study. We find $k_{\Sigma \pi^0} \leq 0.1$ in 10-20\% of the events and $k_{\Sigma \pi^0} \geq 0.9$ below the 0.2\textperthousand{} level. An energy dependence is negligible for the highest $k_{\Sigma \pi^0}$, slight energy dependence for lower $k_{\Sigma \pi^0}$s is visible.

\section{Effect of the $\pi^0$ decay length at the highest energies }\label{lorentz}

\begin{figure*}[t]
\centering
\begin{minipage}{0.48\textwidth}
\centering
\includegraphics[width=1\columnwidth]{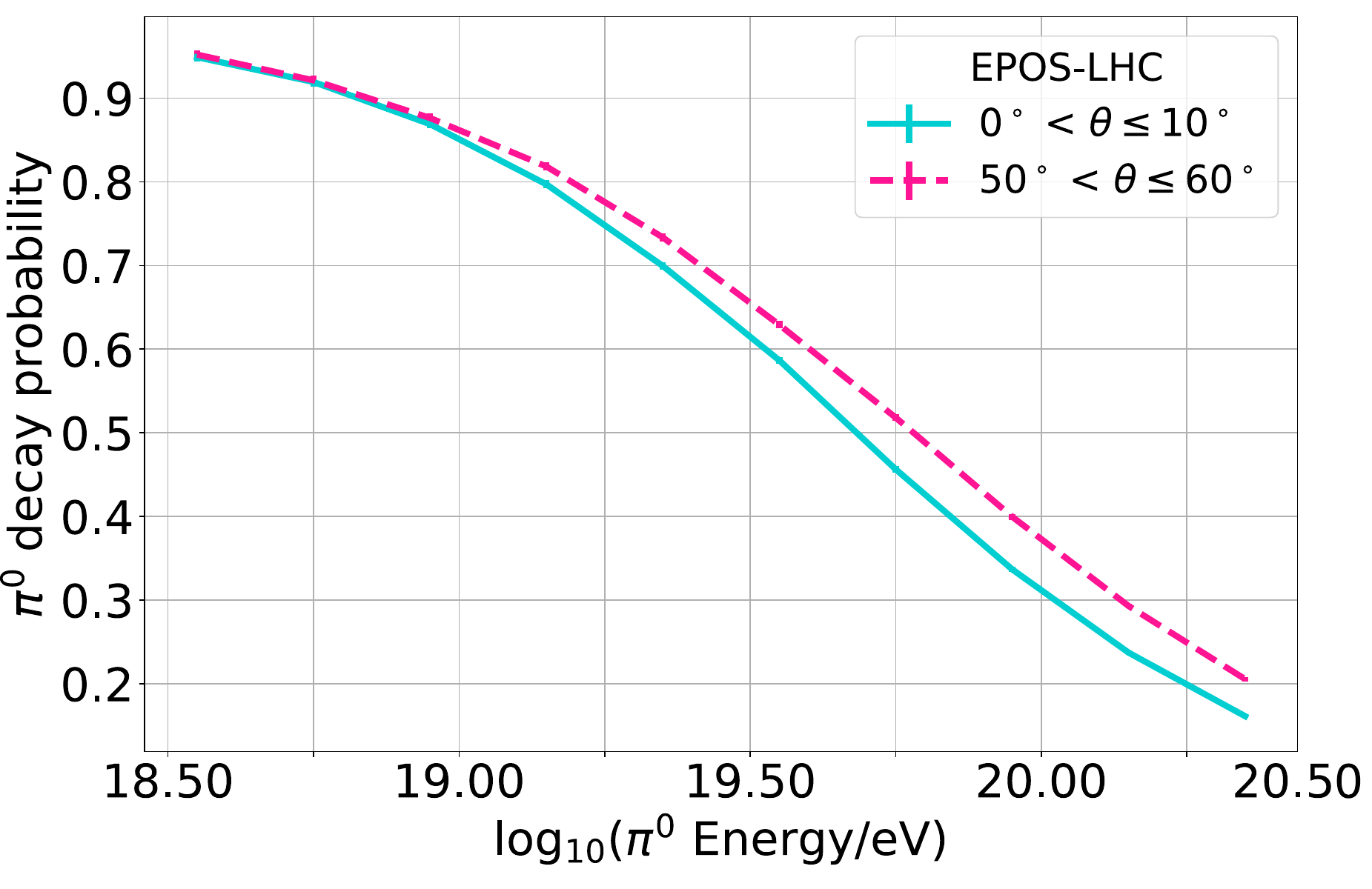}
\caption{Decay probabilities of injected $\pi^0$ as a function of their energy for two different zenith angles (blue: vertical, magenta: inclined) simulated with the EPOS-LHC model.}
\label{decay_probab}
\end{minipage}%
\hspace{5mm}
\begin{minipage}{0.48\textwidth}
\centering
\includegraphics[width=1\columnwidth]{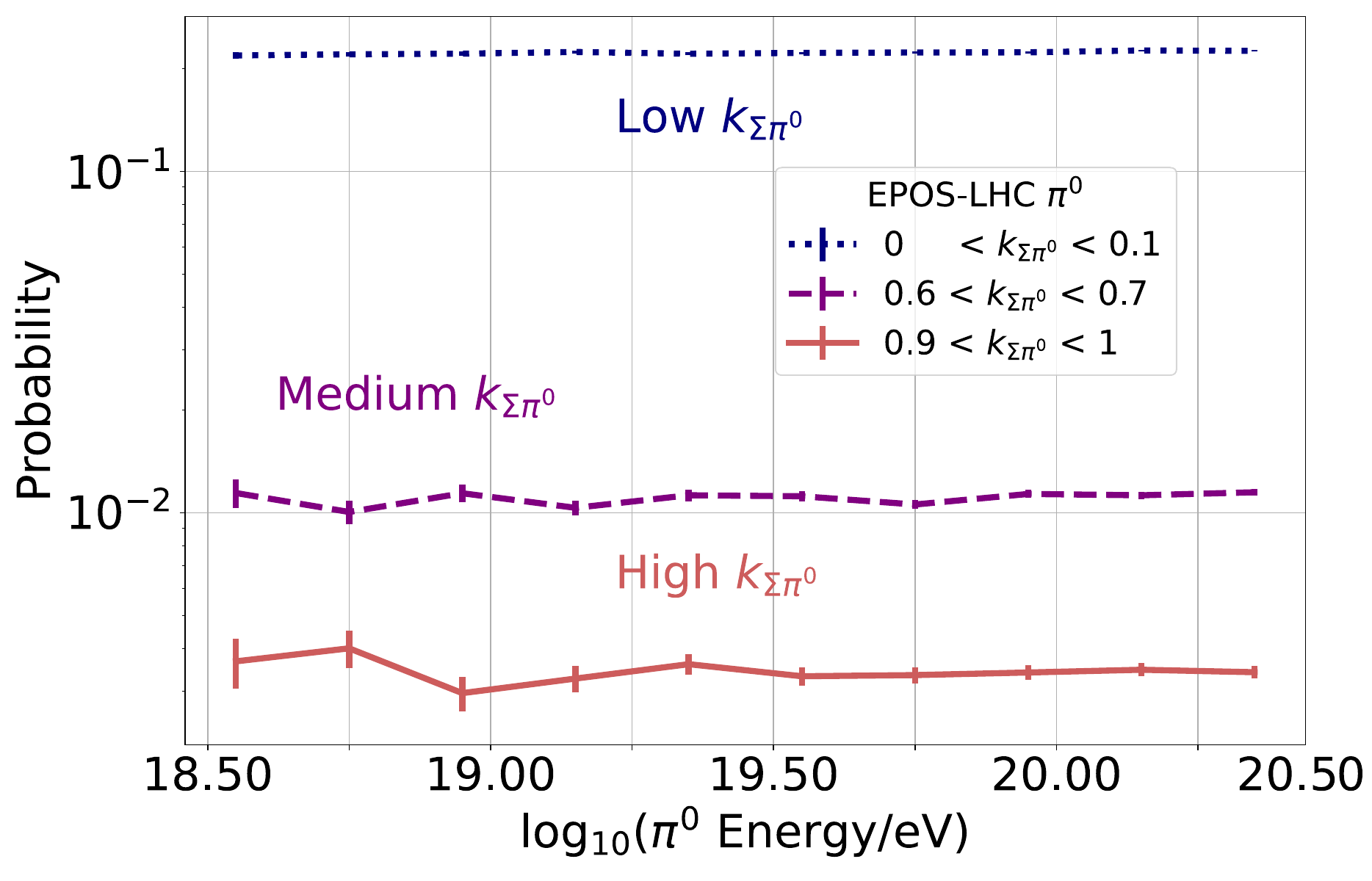}
\caption{Energy dependent probability to find specific ranges of $k_{\Sigma \pi^0}$ (blue: low, purple: medium, red: large) in case of a $\pi^0$-interaction for the EPOS-LHC model.}
\label{repion}
\end{minipage}
\end{figure*}
As discussed above, the observed probabilities of finding large values of $k_{\Sigma \pi^0}$ depends only very weakly on the primary energy. However, the impact that large values of $k_{\Sigma \pi^0}$ can have to the shower development is significant, once the energy of $\pi^0$'s exceed $\sim 1$\,EeV. Above these energies, the $\pi^0$ decay length becomes sufficiently large to allow hadronic interactions taking place before the $\pi^0$ decays into two photons. For example, at $E_{\pi^0}=10$\,EeV 
the decay length $\tau_{\rm decay}=\gamma \, c \, \tau_0$ (with $\gamma$ and $\tau_0$ being the Lorentz factor and lifetime of the $\pi^0$, respectively, and $c$ the speed of light) reaches about 2\,km, which starts to exceed the pion-Air hadronic interaction length at atmospheric heights below about 6\,km.
Therefore, the enhancement of the electromagnetic component of hadron-induced air showers making them more photon-like due to upward fluctuations of $k_{\Sigma \pi^0}$ will be counteracted at the highest energies by re-interactions of $\pi^0$ prior to their decay. The onset of $\pi^0$ re-interactions above cosmic ray energies of $\sim 1$\,EeV has been noted also in \citealt{dEnterria:2018kcz}.

The probability for $\pi^0$'s suffering hadronic interactions dependents mainly on the Lorentz boost and local density of traversed atmosphere, the latter being dependent on the zenith angle of the incoming cosmic ray.
To quantify both effects, we simulated a sample of $10^6$ neutral pions as primary particles. They were injected at heights following the distribution of the observed first interaction heights of the proton simulations. The primary energy range was extended down to $10^{18.5}$\,eV and the zenith angle ranges from $0^\circ$ to 60\,$^\circ$. During their propagation, the decay and hadronic interaction of the $\pi^0$ compete according to their decay length and hadronic cross section.
The resulting decay probabilities are shown in Fig.\ \ref{decay_probab}. For the lower energies, approximately 95\% of the $\pi^0$s  decay, regardless of the zenith angle. At energies above $10^{19.75}$\,eV the hadronic interactions dominate and a modest difference between low and high zenith angles is obtained. This trend continues towards the highest considered energies, at which only 16\% of the $\pi^0$'s in  $\theta \in
[0^\circ, 10^\circ]$ decay. In the range  $\theta \in [50^\circ, 60^\circ]$
we find a decay probability of 21\%, which means a relative difference of 36\%. Nonetheless, it is evident that the energy is the primary determinant of the increased interaction probability.

In case of a hadronic interaction, the $\pi^0$ can again produce high energy $\pi^0$'s. The probabilities of the corresponding $k_{\Sigma \pi^0}$ ranges is depicted in Fig.\ \ref{repion}. Compared to proton primaries of the same energy (\textit {c.f.} Fig.\,\ref{prob_epos}), the probability of finding a large fraction of the incoming energy being transferred to neutral pions, e.g.\ $k_{\Sigma \pi^0} \geq 0.8$, is enhanced by about one order of magnitude. Nevertheless, only a small fraction of events remains photon-like after the hadronic interaction of a $\pi^0$.

\begin{figure*}
\centering
\begin{minipage}{0.4\textwidth}
\centering
\includegraphics[width=1\columnwidth]{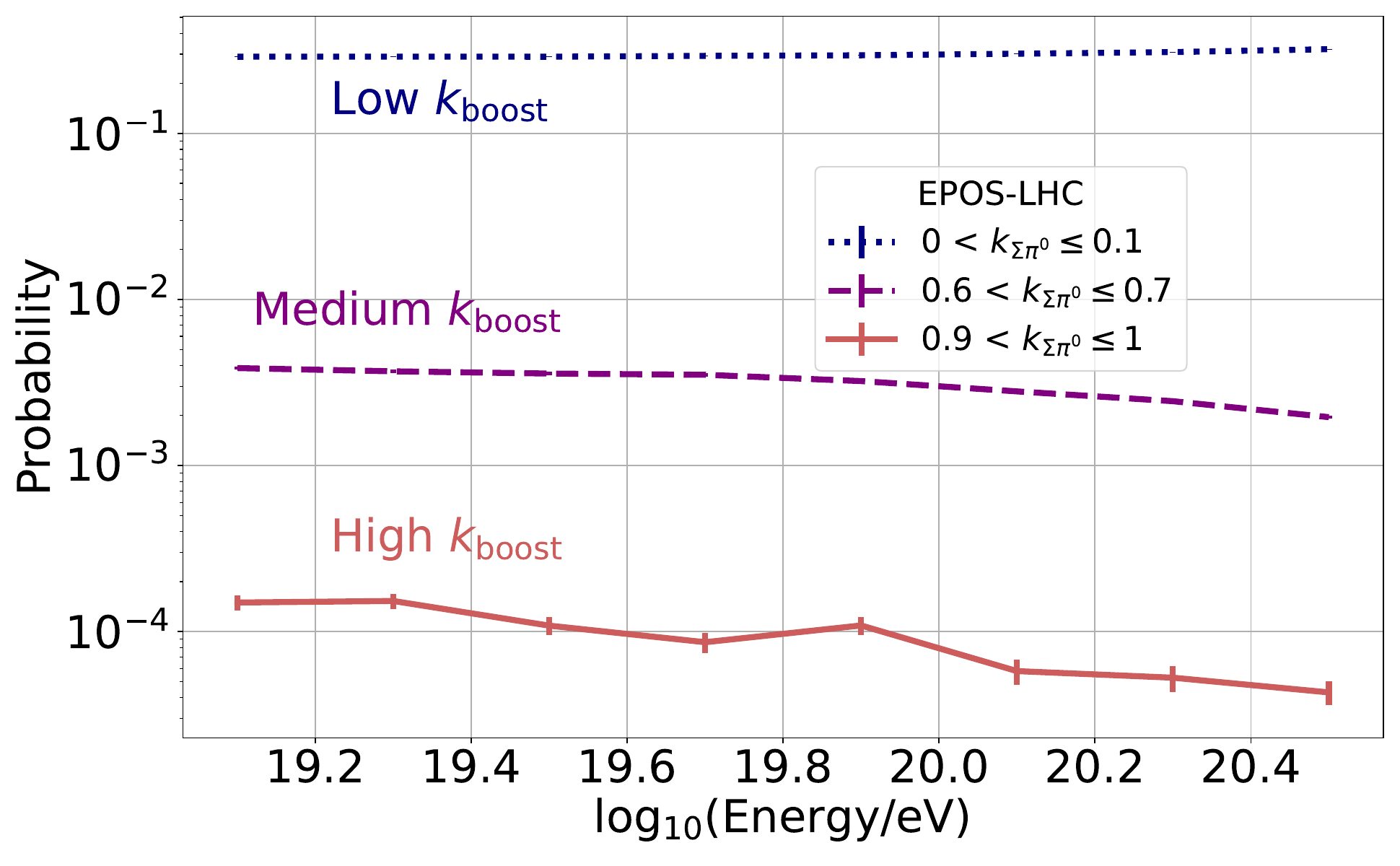}
\caption{Energy dependent probability with Lorentz boosting correction to have specific $k_\text{boost}$ (blue: low, purple: medium, red: large) in case of a proton interaction for the EPOS-LHC model.}
\label{final_prob}
\end{minipage}%
\hspace{5mm}
\begin{minipage}{0.52\textwidth}
\centering
\includegraphics[width=1\columnwidth]{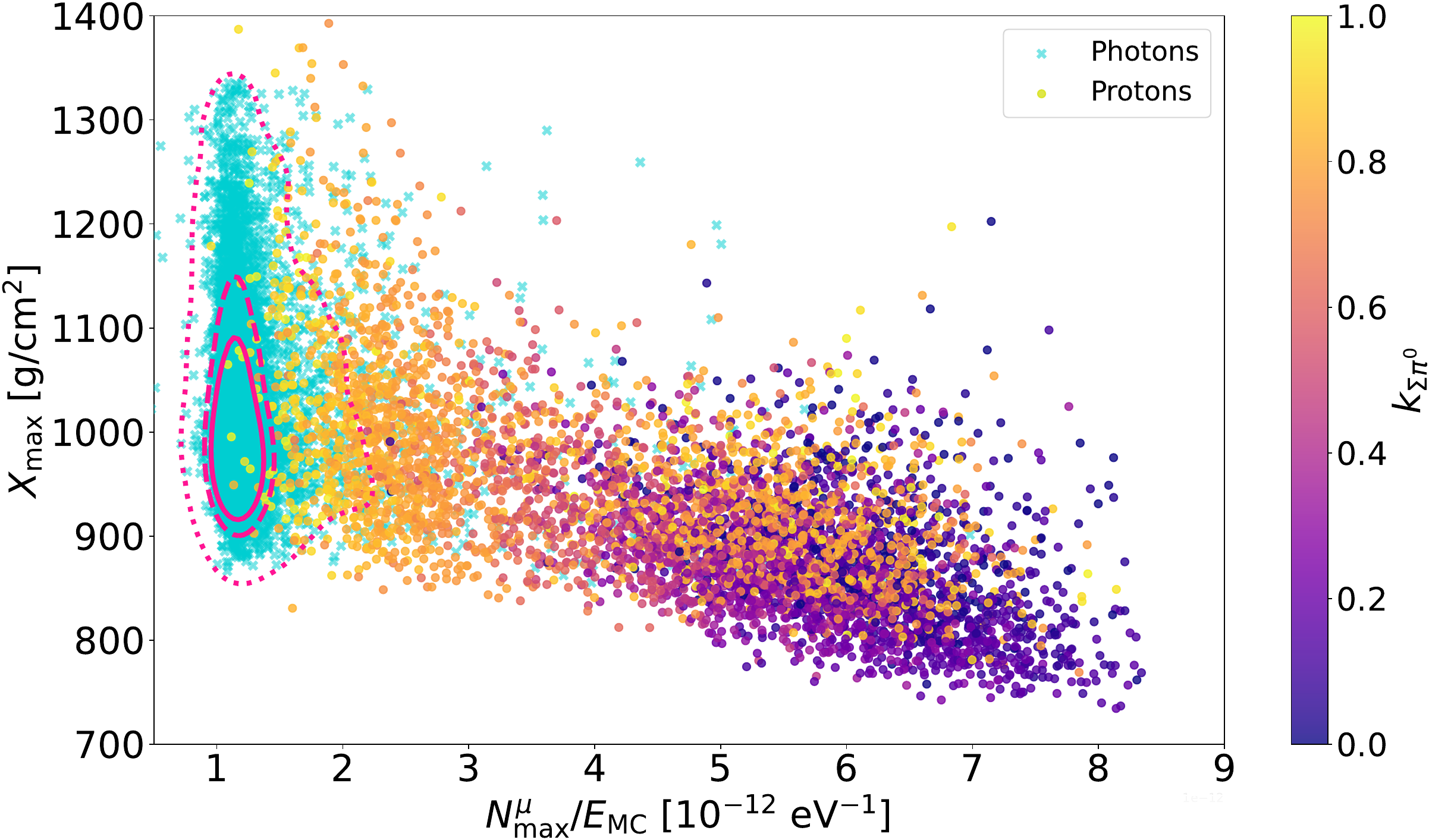}
\caption{Heatmap of the simulated shower maximum $X_\text{max}$ versus the normalized number of muons at shower maximum, $N^\mu_\text{max}$ for proton (colored points) and photon (light blue crosses) induced air showers at primary energies of $E_0 \in [10^{19}, 10^{20.5}]$\,eV. The magenta intensity contours (50\,\%, $1\sigma, 2\sigma$) of the photon induced showers are included for comparison and the $k_\text{boost}$ values of the proton induced showers are color coded.
}
\label{scat_xmax_nmax1}
\end{minipage}
\end{figure*}

Applying the hadronic interaction probabilities to the $\pi^0$'s produced in the first cosmic ray interaction (combining Figs.\ \ref{prob_epos}, \ref{decay_probab} and \ref{repion}), results in Fig.\ \ref{final_prob}. The altered photon-likeness, regarding the Lorentz boosting, is here denoted as $k_\text{boost}$ ($k_\text{boost} := k_{\Sigma \pi^0} - $ losses due to decay following Lorentz boosting). A distinct energy dependence is obtained: for low $k_\text{boost}$, the probability increases with energy by $\approx$ 10\,\%. For high $k_\text{boost}$, a reduction in probability by factors up to four is observed. It is therefore evident that photon-like events are most likely found at lower energies, as also the proton abundance of the primary cosmic ray flux is higher here\,(\citealt{spectral_indices}).

To investigate how well a photon induced air shower can be discriminated from a cosmic ray induced photon-like shower as a function of $k_{\Sigma \pi^0}$, full CORSIKA simulations were performed for about 8,100 photon- and 6,700 proton induced showers at energies $E_0 \in [10^{19}, 10^{20.5}]$\,eV. They were selected to be uniformly distributed across the whole $k_{\Sigma \pi^0}$ range. Figure \ref{scat_xmax_nmax1} shows the main discrimination variables presently used in photon searches: the muon fraction and the depth of the shower maximum, $X_\text{max}$. The muon fraction is estimated in this study as the number at shower maximum, $N^\mu_\text{max}$, extracted from the MC values of the CORSIKA simulations and normalized by the primary energy. The color scale indicates the initial $\pi^0$-content $k_{\Sigma \pi^0}$ of the individual proton showers. Protons induced showers with a low $k_{\Sigma \pi^0}$ content are clearly separated from the photon distribution. With increasing $k_{\Sigma \pi^0}$, the variables shift towards the photon distribution and start to merge with it at the highest $k_{\Sigma \pi^0}$ values. These data points are entirely situated within the photon distribution and cannot be distinguished from photons, even with a fully efficient detector. This background caused by proton primaries is irreducible. Although a significant number of events with high $k_{\Sigma \pi^0}$ values is observed within the $1 \sigma$ and even in the 50\% contour of the photon distribution, the majority of this group of events is found outside the $2\sigma$ photon region ranging even out into the tail where events with low $k_{\Sigma \pi^0}$ values are observed.
These are mostly high energy events with increased probability of the produced $\pi^0$'s to re-interact so that the showers resemble a regular proton showers. Hence, they have a high $k_{\Sigma \pi^0}$ but a low $k_\text{boost}$.

\section{Expectations for the irreducible background}\label{chap4}

\begin{figure*}[tbh]
\centering
\includegraphics[width=0.86\textwidth]{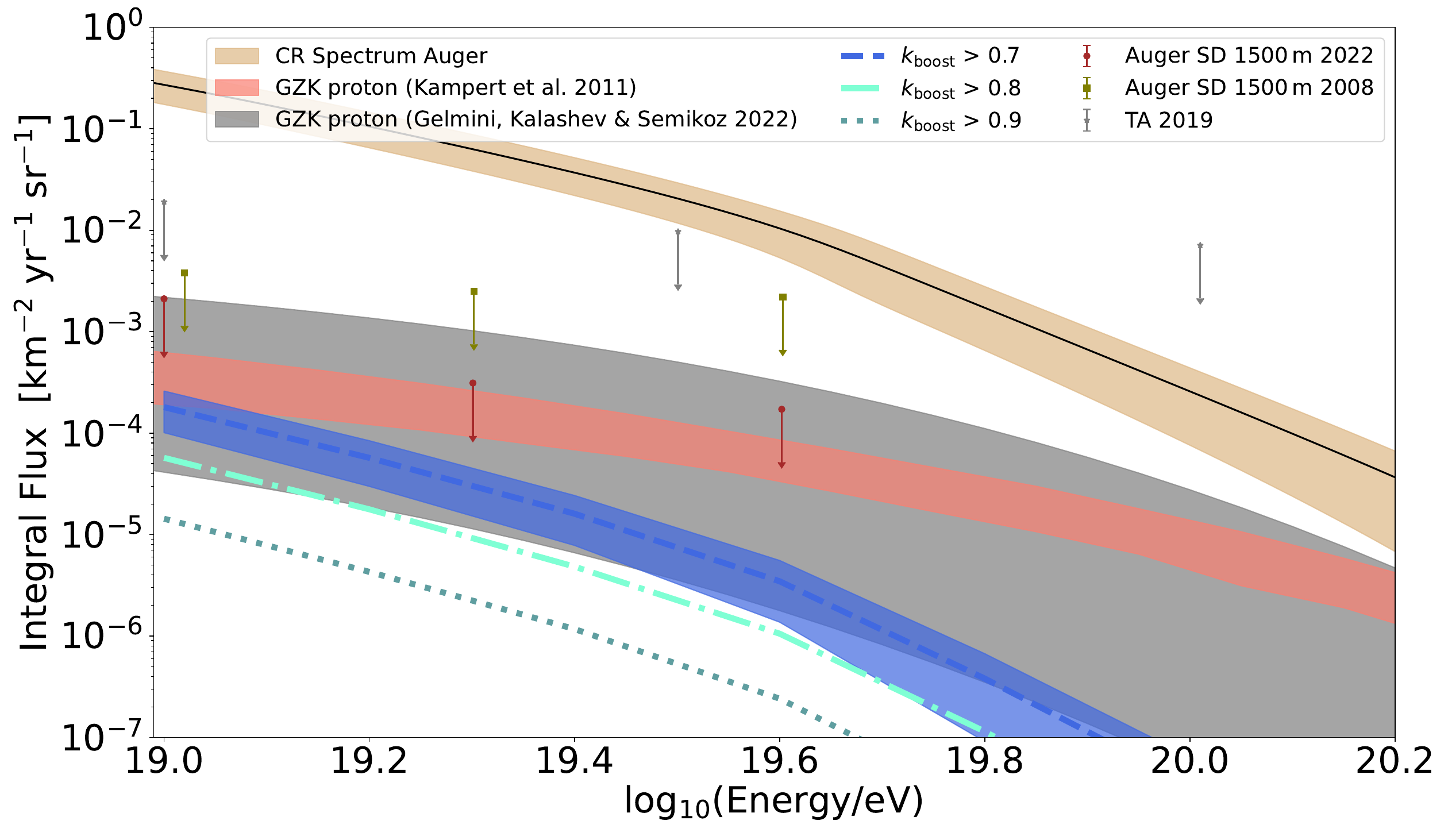}
\caption{The integral fluxes of high $k_\text{boost}$ events based on the spectrum measured by the Pierre Auger Observatory together with the flux limits from \citealt{PierreAuger:2022aty,sd_limits,ta_limits} are shown. Flux predictions from the GZK effect for a pure proton spectrum are shown as grey and red bands. For reasons of clarity, the statistical uncertainty of the photon-like background is only shown for $k_\text{boost} > 0.7$, as the other bands are consistent with~0. The plot shows a consistency of the photon limits with the irreducible proton background.}
\label{flux_final}
\end{figure*}

We now compare the flux of proton induced air showers of different $k_\text{boost}$ with recent upper bounds on the photon flux, shown in Fig.\ \ref{flux_final}. The upper (light brown) band represents the all-particle cosmic ray spectrum published in Ref.\ \citealt{spectral_indices}. The different dashed and dash-doted lines are derived from the cosmic ray  flux after accounting for the extracted $k_\text{boost}$ regions of these events. These curves represent the expectations for the irreducible hadronic background. As can be seen, the background distributions remain well below the published upper limits on UHE photons (\citealt{PierreAuger:2022aty,sd_limits,ta_limits}) for $k_\text{boost} > 0.7$ and are approximately one order of magnitude lower at 10\,EeV with the difference increasing towards higher energies. 
We also compare the expected flux of photon-like background events with the flux of cosmogenic photons expected from the GZK effect (assuming again a pure proton spectrum).
This indicates that the expected background of photon-like events enters into the GZK predictions (\citealt{kampert:2011gzk, gelmini:2022gzk}) for $k_\text{boost} > 0.7$ and remains below only for $k_\text{boost} > 0.8$.

The consistency of the simulations with the published photon bounds is an interesting result of this analysis. However, more intriguing is the comparison between the anticipated photon-like proton-induced air showers and the number of observed photon candidates. For a quantitative study, we will work with the exposure $A = 46,900$\,km$^2$\,sr\,yr from Ref.\ \citealt{PierreAuger:2022aty} to which the Monte Carlo simulations will be normalised. For a most conservative estimate of the number of photon-like events, we will assume a pure proton energy spectrum. The spectrum is taken from the Pierre Auger Collaboration\,(\citealt{spectral_indices}). The integrated spectrum is then multiplied with the exposure and with the probabilities to find events with specific values of $k_\text{boost}$. This yields the number of expected background events $N_\text{bg}$ summarized in Table\,\ref{candidates}.

\begin{table}[!h]
\centering
\caption{Number of expected background events for different $k_{\Sigma \pi^0}$ ranges and energies above 10\,EeV.}
\label{candidates}
\begin{tabular}{|c|c|c|c|c|}
\hline
$k_\text{boost}$ & 0.6 - 1   & 0.7 - 1     & 0.8 - 1       & 0.9 - 1       \\ \hline
$N_\text{bg}$ & 246$\pm$88 & 8.5$\pm$3.7 & 2.7$\pm$2.4 & 0.7$\pm$2.2 \\ \hline
\end{tabular}
\end{table}

One expects $246 \pm 88$ background events for $k_\text{boost} \geq 0.6$ rapidly shrinking down to $8.5 \pm 3.7$ for $k_\text{boost} \geq 0.7$. Hence, only about half of the 16 observed photon candidates reported in \citealt{PierreAuger:2022aty} can be explained by proton induced showers with $k_\text{boost} \geq 0.7$ and only $(16.7 \pm 14.8)$\% by protons with $k_\text{boost} \geq 0.8$. 
The number of expected photon-like background events reduces further when taking into account that the cosmic ray flux at energies above 10\,EeV is not composed of protons only, but comprises a significant fraction of intermediate mass nuclei of $2 \leq Z \leq  22$ increasing even further towards heavier primaries as the energy increases (\citealt{mass_comp2}). Yet another reduction of irreducible background events could be expected from the so-called muon-puzzle (\citealt{Albrecht:2021cxw}), a deficit of muons in simulations relative to air shower data (\citealt{muon_deficit2,muon_deficit}). Enhancing the hadronic component in simulations could lead to a better discrimination between photons and protons and reduce the expected number of irreducible background events. Finally, we remind the reader that EPOS-LHC yields the highest probability for $k_\text{boost}$ events, \textit{c.f.} Fig.\,\ref{distr_all}.

It should be noted, on the other hand, that even modern air shower observatories, such as the Pierre Auger Observatory (\cite{PierreAuger:2015eyc}), will not achieve the ideal discrimination power between photons and hadrons shown in Fig.~\ref{scat_xmax_nmax1}. Considering realistic air shower detection and reconstruction efficiencies, will inevitably increase the photon-like background, particularly when accounting for the much more abundant cosmic ray flux w.r.t.\ photons. Moreover, the energy reconstruction ability of photon and photon-like air showers could play an additional role in estimating the background. It is beyond the scope of the present work to quantitatively analyze these effects.
However, considering only the simulations, even hadron events with $k_\text{boost}$ as low as 0.6 would need to be classified incorrectly as photons, which seems improbable according to Fig.~\ref{scat_xmax_nmax1}. Future studies may evaluate this discrimination capability for the background induced by protons.

\section{Summary and conclusion}\label{summary}

In this paper, the irreducible hadron background expected in searches for UHE photons at $E_0 > 10$\,EeV has been studied and it was demonstrated that an upward fluctuation in the production of high energy $\pi^0$'s in the first cosmic ray interactions in the atmosphere can mimic the properties of photon induced air shower. This may either be due to a single high energy $\pi^0$ produced in the first interaction or due to several $\pi^0$'s carrying away a significant portion of the initial cosmic ray energy, both being quantified by the relative energy share $k_{\Sigma \pi^0}$.  For a quantitative analysis, two hadronic interaction models were considered: EPOS-LHC and QGSJetII-04. It is found that the probability of finding events with large values of $k_{\Sigma \pi^0}$ is significantly higher for EPOS-LHC than for QGSJetII-04. In principle, more hadronic interaction models could be considered (e.g. Sibyll 2.3d (\cite{Riehn_2020})). Newer models like the EPOS LHC-R (\cite{PierogWerner2023_1000163321}) could contribute to a more precise estimate of the irreducible proton background and might show more interesting features. However, for this study, EPOS-LHC can be used for a conservative background estimate.

The probability for photon-like events is found to depend not only on the produced secondaries, but to vary also strongly with the primary energy for $E_0 \gtrsim 10^{18.5}$\,eV. This is because at these energies the decay length of $\pi^0$'s starts to become comparable or even exceed their hadronic interaction length in the atmosphere. Thus, the number of cosmic ray induced photon-like events sharply decreases at the highest energies. It is also shown that heavier nuclei as primaries like helium have no significant contribution to the number of photon-like air showers.

By combining the calculated probabilities for finding different values of $k_\text{boost}$ with the actually measured number of cosmic ray events as a function of energy, we find inconsistencies between the events found in the data sample and those observed as background $N_\text{bg}$ in \cite{PierreAuger:2022aty}. The significance of this inconsistency depends on the benchmark $k_\text{boost}$, which needs to be determined by the  discrimination power of the respective analysis.

We note that the background estimates presented here are conservative in a number of respects so that  theoretically expected hadron induced background is likely much lower, and could be increased only due to detector resolution and event reconstruction effects. However, based on our present understanding we conclude that the proton background caused by high energy $\pi^0$'s in the first interaction is unlikely to serve as an explanation for the identified photon candidates. Nonetheless, integrating photon-like air shower simulations into photon searches (e.g. by the Pierre Auger Observatory) could yield valuable insights on how well the discrimination analysis performs. Vice versa, measured data could help to validate the findings of this study on the maximum irreducible proton background.

\subsection*{Acknowledgments} We thank the members of the Pierre Auger Collaboration for fruitful discussions related to this topic.
The simulations were carried out on the pleiades cluster at the University of Wuppertal, which was supported by the Deutsche Forschungsgemeinschaft (DFG).
Further financial support has been provided by the Bundesministerium für Bildung und Forschung (BMBF-Verbundforschung).

\bibliography{main.bib} 

\end{document}